\documentclass[a4paper,fleqn]{cas-dc}

\usepackage[numbers]{natbib}

\usepackage{multirow}
\usepackage{hyperref}
\usepackage{tabularx}
\usepackage{textcomp}
\usepackage[ruled,linesnumbered]{algorithm2e}

\def\tsc#1{\csdef{#1}{\textsc{\lowercase{#1}}\xspace}}
\tsc{WGM}
\tsc{QE}
\tsc{EP}
\tsc{PMS}
\tsc{BEC}
\tsc{DE}

\begin{document}
\let\WriteBookmarks\relax
\def\floatpagepagefraction{1}
\def\textpagefraction{.001}

\shorttitle{Vessel centerline extraction}

\shortauthors{Sijie Liu et~al.}

\title [mode = title]{An automated framework for brain vessel centerline extraction from CTA images}                      




%

\author[1,2]{Sijie Liu}[type=editor]


\fnmark[1]



\credit{Conceptualization, Methodology, Software, Writing - original draft}

\affiliation[1]{organization={Institute of Artificial Intelligence and Robotics, Xi'an Jiaotong University, Xi'an, China}}
\cormark[1]
\ead{liusijie@stu.xjtu.edu.cn}

\author[2]{Ruisheng Su}
\affiliation[2]{organization={Department of Radiology, Erasmus MC, Rotterdam, The Nethelands},
}
\fnmark[1]
\credit{Conceptualization, Methodology, Writing - original draft}
    
\author[2]{Jiahang Su}

\credit{Investigation, Data curation, Software}

\author[1]{Jingmin Xin}
\credit{Supervision, Writing - review \& editing}

\author[1]{Jiayi Wu}
\credit{Writing - review \& editing}

\author[3]{Wim van Zwam}
\affiliation[3]{organization={Department of Radiology, Maastricht UMC, Maastricht, The Netherlands},
}
\credit{Data curation, Writing - review \& editing}

\author[2]{Pieter Jan van Doormaal}
\credit{Data curation, Writing - review \& editing}

\author[2]{Aad van der Lugt}
\credit{Writing - review \& editing}

\author[2,4]{Wiro J. Niessen}
\affiliation[4]{organization={Faculty of Applied Sciences, Delft University of Technology, Delft, The Netherlands},
}
\credit{Writing - review \& editing}

\author[1]{Nanning Zheng}
\credit{Funding acquisition, Writing - review \& editing}

\author[2]{Theo van Walsum}
\credit{Funding acquisition, Supervision, Software, Writing - review \& editing}

\cortext[cor1]{Corresponding author.}

\fntext[fn1]{Sijie Liu and Ruisheng Su contributed equally to this work.}


\begin{abstract}
Accurate automated extraction of brain vessel centerlines from CTA images plays an important role in diagnosis and therapy of cerebrovascular diseases, such as stroke.
However, this task remains challenging due to the complex cerebrovascular structure, the varying imaging quality, and vessel pathology effects.
In this paper, we consider automatic lumen segmentation generation  without additional annotation effort by physicians and more effective use of the generated lumen segmentation for improved centerline extraction performance.
We propose an automated framework for brain vessel centerline extraction from CTA images.
The framework consists of four major components:
(1) pre-processing approaches that register CTA images with a CT atlas 
and divide these images into input patches, 
(2) lumen segmentation generation from annotated vessel centerlines 
using graph cuts and robust kernel regression, 
(3) a dual-branch topology-aware UNet (DTUNet) that can effectively utilize the annotated vessel centerlines and the generated lumen segmentation through a topology-aware loss (TAL) and its dual-branch design,
and (4) post-processing approaches that skeletonize the predicted lumen segmentation.
Extensive experiments on a multi-center dataset demonstrate that the proposed framework 
outperforms state-of-the-art methods in terms of average symmetric centerline distance (ASCD) 
and overlap (OV). 
Subgroup analyses further suggest that the proposed framework holds promise in clinical applications for stroke treatment.
Code is publicly available at \url{https://github.com/Liusj-gh/DTUNet}.
\end{abstract}


\begin{highlights}
\item An automated framework is proposed for brain vessel centerline extraction from CTA images.
\item A lumen segmentation generation algorithm based on annotated centerlines is used to provide additional supervision information during the training phase.
\item A dual-branch topology-aware UNet is used to effectively utilize the annotated vessel centerlines and the generated lumen segmenation through a topology-aware loss and its dual-branch design.
\end{highlights}


\maketitle

\section{Introduction}
Stroke is globally the second leading cause of death and the third leading cause of disability, with approximately 15 million people each year suffering from a stroke worldwide~\cite{who}.
As an acute cerebrovascular disease, stroke is featured by brain vessel changes (e.g., rupture and occlusion).
Therefore, the cerebrovascular structure is relevant in the clinical setting for prevention, assessment, prognosis, and treatment of stroke~\cite{peisker2017}.
Computed tomographic angiography (CTA) is a minimally invasive imaging technique that can visualize intracranial vessels and is widely used for diagnosis and treatment decision-making of stroke patients.
Automated analysis of intracranial vessels and detection of related diseases could provide valuable support to clinicians and interventionists~\cite{el-Baz2012,soltanpour2021}. For example,
cerebrovascular structure extraction can be used to localize vessel occlusion~\cite{stib2020} and calculate collateral scores~\cite{su2020}. Automated centerline extraction can also assist in vessel segmentation~\cite{lv2019}, brain registration~\cite{reinertsen2007}, etc. 
However, accurate automated extraction of brain vessel centerlines from CTA images is still a challenging task due to the complex cerebrovascular structure, the varying imaging quality, and vessel pathology effects.

\begin{figure*}[t]
	\centering
	\includegraphics[width=0.92\linewidth]{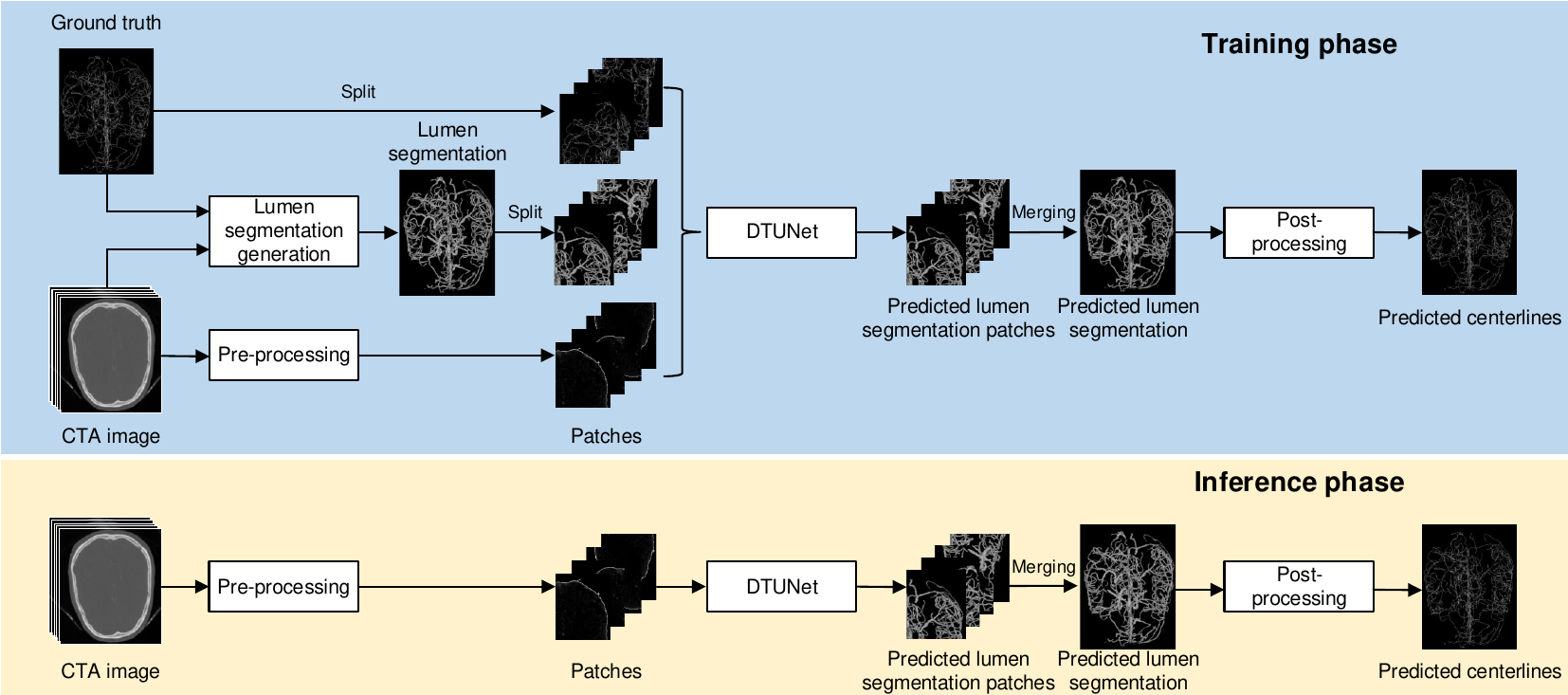}
	\caption{An overview of the proposed framework. During the training phase, it consists of four major components, including pre-processing approaches, lumen segmentation generation, DTUNet, and post-processing approaches. \textbf{Note that the proposed framework does not require the lumen segmentation generation during the inference phase.}}
	\label{fig:pipeline}
\end{figure*}

\subsection{Traditional methods for vessel centerline extraction}
Traditional methods for vessel centerline extraction can be divided into three categories:
morphology-based, distance transformation-based, and model evolution-based methods. 
Morphology-based methods skeletonize pre-segmented 3D vessels to obtain centerlines.
Pal{\'a}gyi et al.~\cite{palagyi2001} proposed an 8-direction thinning rule to extract vessel topologies.
Cheng et al.~\cite{cheng2014} proposed a robust thinning algorithm that used line direction filters in thirteen orientations to enhance small vessels.
Distance transformation-based methods extract centerlines from a distance map 
where each voxel value represents the minimum distance from that voxel to vascular edges.
He et al.~\cite{he2001} used a reliable path approach to find central paths 
which could be subsequently converted to vessel centerlines.
Jin et al.~\cite{jin2016} proposed a curve skeletonization algorithm based on a minimum cost path.
Model evolution-based methods obtain initial centerlines using axis detection, 
and then evolve the initial centerlines to correct central positions by minimizing an energy function.
Krissian at al.~\cite{krissian2000} extracted a centerline and estimated its radius simultaneously by a multiscale centerline detection algorithm based on a cylindrical model.
Fung et al.~\cite{fung2013} modeled vessel centerlines as cubic B-spline curves composed of 3D cubic polynomial segments.

\subsection{Deep learning methods for vessel centerline extraction}
Recently, deep learning methods have achieved promising results in vessel segmentation by virtue of their powerful feature representation capabilities~\cite{moccia2018,xiao2023}.
Therefore, some studies have leveraged lumen segmentation results to extract vessel centerlines.
Guo et al.~\cite{guo2019} employed a multi-task fully convolutional network to generate centerline distance maps and
endpoint maps from segmentation masks of coronary CTA images. They then applied a minimal cost path algorithm to extract centerlines based on the centerline distance maps and the endpoint maps.
Tetteh et al.~\cite{tetteh2020} presented a DeepVesselNet which could handle multiple vessel-related tasks including vessel segmentation, centerline extraction, 
and bifurcation detection. 
However, the DeepVesselNet relies on the outcome of the vessel segmentation to train the centerline extraction model.
He et al.~\cite{he2020} used vessel segmentation to guide centerline heatmap regression, followed by a minimal cost path to obtain the final vessel centerline.
These works rely on the extra annotation information of the lumen segmentation to extract the vessel centerline.
However, since the lumen in a brain exhibits significant variations in shape, size, and intensity, 
it is usually challenging and time-consuming to precisely determine the lumen border of massive cerebral vessels. Besides, current deep learning methods often require a large amount of training data, which further increases the labor burden on experienced annotators.

There are also attempts that only rely on the annotation of vessel centerlines.
Sironi et al.~\cite{sironi2015} proposed a multiscale centerline regressor to predict distance maps generated from annotated centerlines, 
and used a non-maximum suppression algorithm to obtain the centerlines from the predicted distance maps.
Rjiba et al.~\cite{rjiba2020} proposed a CenterlineNet to extract the main and side branches of coronary artery centerlines.
Su et al.~\cite{su2020} used a 3D UNet to directly extract dilated vessel centerlines from brain CTA images.
Although these studies show promising results, they do not consider topological connections among different vessel segments.
Zhang et al.~\cite{zhang2022} designed a C-UNet for preliminary vessel centerlines extraction from coronary angiography images. They then used a multifactor centerline reconnection algorithm to improve the continuity of the extracted centerlines.
Some studies~\cite{zhang2018, wolterink2019} investigated tracking-based approaches to extract vessel centerlines. While these approaches are well-suited for simple scenarios with a few vessel centerlines, they may be less suitable for brains with hundreds of vessel centerlines due to increased complexity.

Besides, a variety of loss functions have been proposed to improve the performance of vessel centerline extraction.
Wu et al.~\cite{wu2020} proposed a voxel-based centerline loss to guide an asymmetric encoding and decoding-based convolutional neural network to allocate equal attention to small vessels and large vessels in brain CTA images.
Dorobantiu et al.~\cite{dorobantiu2021} also utilized a 3D UNet optimized by an adapted loss function to address a sparse annotation issue for coronary centerline extraction from cardiac CTA images.
Shit et al.~\cite{shit2021} proposed a novel centerline Dice (clDice) loss function to guarantee topology preservation for binary 2D and 3D segmentation.
Pan et al.~\cite{pan2021} further applied the clDice loss function to simultaneously improve centerline extraction and vessel segmentation of 2D retinal fundus images.
However, the number and complexity of blood vessels in 3D brain CTA images are considerably higher than those in 2D retinal fundus images. 

\subsection{Contributions}
In this paper, we consider automatic lumen segmentation generation without additional annotation effort by physicians and more effective use of the generated lumen segmentation for improved centerline extraction performance.
To this end, we propose an automated framework for brain vessel centerline extraction from CTA images.
The framework not only generates lumen segmentation based on annotated vessel centerlines, but also effectively exploits both the annotated vessel centerlines and the generated lumen segmentation during the training phase using a dual-branch topology-aware UNet (DTUNet).
Specifically, the framework, as shown in Fig.~\ref{fig:pipeline}, consists of four major components: (1) pre-processing approaches that register CTA images with
a CT atlas to mask the brain, and split these images into input patches, 
(2) lumen segmentation generation from annotated vessel centerlines using graph cuts and robust kernel regression, 
(3) a DTUNet that can effectively utilize the annotated vessel centerlines and the generated lumen segmentation through a topology-aware loss (TAL) and its dual-branch design,
and (4) post-processing approaches that skeletonize the predicted lumen segmentation.
A multi-center dataset including 10 intracranial CTA images and 40 cube CTA images is used to evaluate the proposed framework.
In summary, the main contributions are as follows:
\begin{itemize}
	\item We demonstrate that introducing the lumen segmentation generated from annotated centerlines during the training phase can improve the performance of brain vessel centerline extraction. Notably, during the inference phase, the lumen segmentation generation is no longer required, and thus it does not increase the computational overhead.
	\item We propose a DTUNet. During the training phase, it not only effectively exploits both the annotated vessel centerlines and the generated lumen segmentation in a multi-task learning manner, but also ensure the consistency of the topological connectivity between the vessel centerlines and the lumen segmentation using a TAL.
	\item We demonstrate the superior performance and clinical potential of the proposed framework through comprehensive experiments and subgroup analyses.
\end{itemize}

\section{Method}
The proposed framework for brain vessel centerline extraction from CTA images is illustrated in Fig.~\ref{fig:pipeline}.
It consists of four major components: (1) pre-processing,
(2) lumen segmentation generation, (3) DTUNet, and (4) post-processing.
Note that the proposed framework does not require lumen segmentation generation during the inference phase.
The details of all the  components are elaborated in the following.

\subsection{Pre-processing}
\begin{figure}[t]
	\centering
	\includegraphics[width=0.92\linewidth]{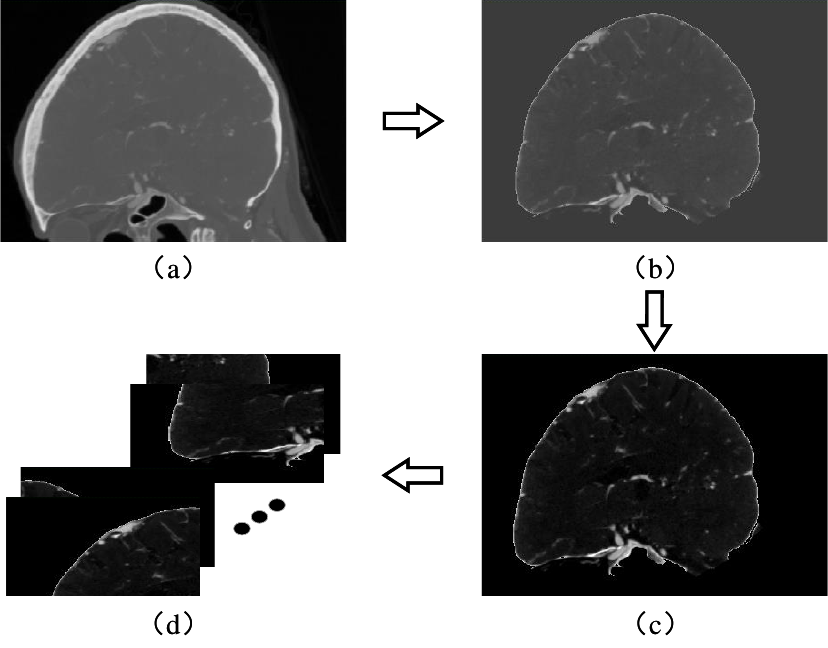}
	\caption{An illustration of the pre-processing approaches. (a) A raw CTA image;
		(b) an image after registration and masking;
		(c) a clipped and normalized image; (d) split patches.}
	\label{fig:pre_processing}
\end{figure}

The purpose of the pre-processing approaches is to align CTA images with a CT atlas 
and to subdivide the CTA images into patches. We adopt similar pre-processing approaches as in~\cite{su2020}, as shown in Fig.~\ref{fig:pre_processing}.
First, we use a CT altas generated from averaging high-resolution 3D CTA images 
of 30 healthy subjects~\cite{friston2003} 
as well as its corresponding binary brain mask.
Subsequently, we use advanced normalization tools (ANTs)~\cite{avants2009}
to register the atlas to each CTA image, after which the corresponding registered brain mask is applied to the images to set all non-brain voxels to zero.
Next, the voxel values are clipped between 0 and 850 Hounsfield unit (HU) based on the intensity distribution of annotated vessel centerlines, after which these values are normalized to a range of 0 -- 1.
Lastly, we generate patches from two types of annotated CTA images: CTA images with all intracranial vessel centerlines annotated (named 'intracranial CTA'), and CTA images where vessel centerlines within a cube with $128\times128\times128$ voxels are annotated (named 'cube CTA').  
During the training phase, we split the intracranial CTA images into $64\times 64\times 64$
overlapping patches with a stride of 16 and remove the patches with no vessel present.
From each of the CTA images, we select the 500 patches with the highest vessel centerline densities, and randomly extract 500 patches from the remaining patches.
For the cube CTA images, we split these CTA images into $64\times 64\times 64$ overlapping patches 
with a stride of 32, and remove the patches with no vessel present.
During the inference phase, we split all the CTA images into $64\times 64\times 64$ non-overlapping patches.
Statistics of all the CTA patches are shown in Table~\ref{table:stastics}.

\begin{table}[t]
	\centering
	\caption{Statistics of the CTA patches}
	\begin{tabular}{cccc}
		\toprule
		Phase &Intracranial &Cube  &Total  \\
		\midrule
		Training &10000 &1058 &11058 \\
		Inference &4096 &320 &4416 \\
		\bottomrule
	\end{tabular}
	\label{table:stastics}
\end{table}

\subsection{Lumen segmentation generation}
\begin{figure}[t]
	\centering
	\includegraphics[width=0.92\linewidth]{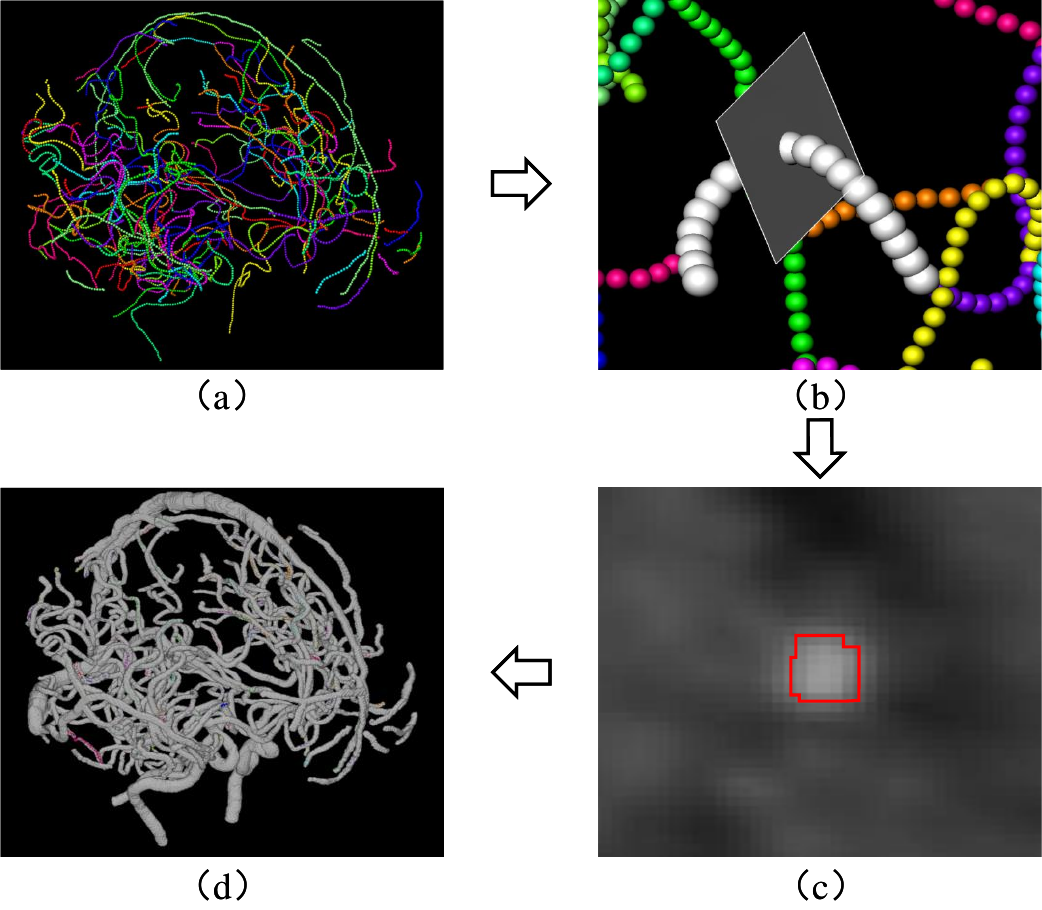}
	\caption{An illustration of the lumen segmentation generation. (a) An annotation of cerebral vessel centerlines where each centerline segment is assigned a different color;
		(b) a cross-sectional slice obtained along a centerline;
		(c) lumen segmentation obtained from a slice using graph cuts and robust kernel regression;
		(d) complete brain lumen segmentation created using an interpolation approach.}
	\label{fig:lumen_generation}
\end{figure}

The purpose of the lumen segmentation generation is to obtain lumen segmentation from annotated centerlines, which can provide additional supervision information for deep neural networks during the training phase.
The lumen segmentation generation consists of three steps, as illustrated in Fig.~\ref{fig:lumen_generation}.
Firstly, following the approach of~\cite{schaap2009}, we resample the image along the annotated centerline~\cite{su2020} to obtain slices orthogonal to the centerline. 
Subsequently, a graph-cut method~\cite{boykov1998} is applied to create lumen segmentation in the cross-sectional slices, followed by robust kernel regression~\cite{debruyne2008} to smooth the border of the lumen segmentation.
Finally, the complete brain lumen segmentation is created from the cross-sectional segmentation using an interpolation approach~\cite{heckel2011}. 
For better understanding, the pseudo-code of the lumen segmentation generation is presented in Algorithm~\ref{alg:lumen_generation}.
During the training phase, similar to the CTA image, the annotated centerline and the generated lumen segmentation are partitioned into patches.
Notably, the annotated centerline and the generated lumen segmentation are no longer required during the inference phase.

\begin{algorithm}[t]
\caption{Lumen segmentation generation}
	\label{alg:lumen_generation}
	\KwIn{Annotated vessel centerlines $V$}
	\KwOut{Brain lumen segmentation $L$}
	$k$ $\gets$ obtain the number of centerline segments in $V$; \\
	\For{$i\gets 1$ \KwTo $k$}{
		$n$ $\gets$ obtain the number of centerline points in the centerline segment $V_{i}$; \\
		\For{$j \gets 1$ \KwTo $n$}{
			$C_{i}^{j}$ $\gets$ obtain the cross sectional slice of the vessel centerline point $V_{i}^{j}$; \\
			$S_{i}^{j}$ $\gets$ predict the lumen segmentation in $C_{i}^{j}$ using the graph-cut method; \\
		}
		$S_{i}$ $\gets$ smooth the border of $S_{i}$ using the robust kernel regression; \\
	}
	\textbf{return} $L$ $\gets$ create the complete brain lumen segmentation from all the lumen segmentation $S$ using the interpolation approach; \\ 
\end{algorithm}

\subsection{DTUNet}
We propose a dual-branch network architecture based on 3D UNet~\cite{cciccek2016} to effectively exploit 
the annotated vessel centerline and the generated lumen segmentation during the training phase. In addition, we utilize a topology-aware loss (TAL) function that ensures 
the consistency of the topological connectivity between vessel centerlines and lumen segmentation. 

\subsubsection{Network architecture}

\begin{figure*}[t]
	\centering
	\includegraphics[width=0.92\linewidth]{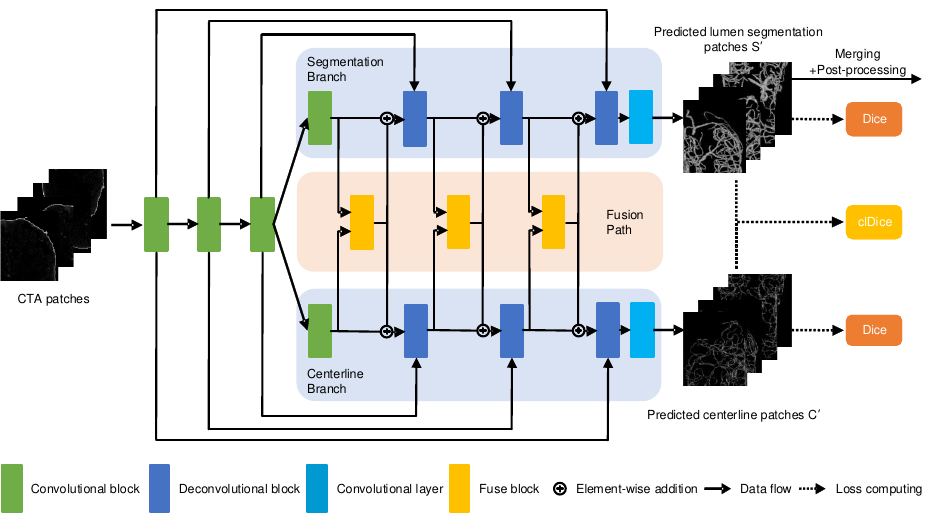}
	\caption{A diagram of the dual-branch topology-aware UNet (DTUNet). \textbf{Note that the predicted lumen segmentation patches $S'$ output by the segmentation branch are used to generate final vessel centerlines via a merging operation and post-processing approaches.}}
	\label{fig:DTUNet}
\end{figure*}

\begin{figure}[t]
	\centering
	\includegraphics[width=0.92\linewidth]{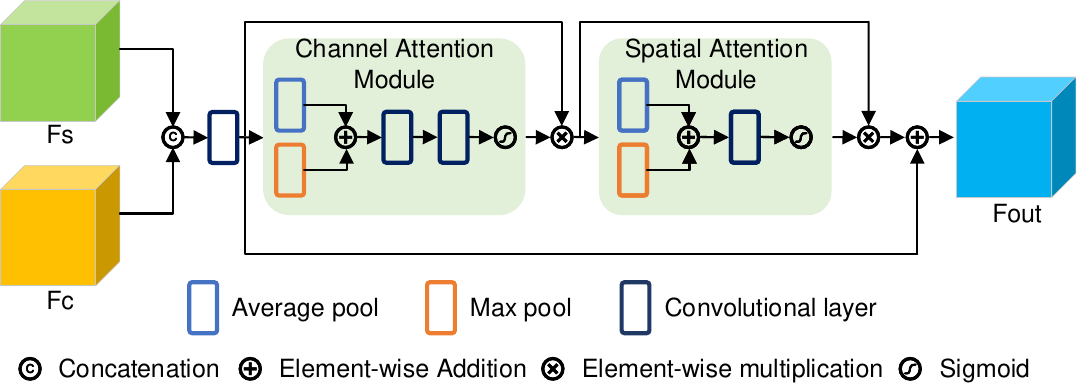}
	\caption{A diagram of the fusion block. It is used to effectively aggregate the feature $F_{s}$ from the segmentation branch and the feature $F_{c}$ from the centerline branch.}
	\label{fig:fusion_path}
\end{figure}

The dual branch topology-aware UNet (DTUNet) is a 3D UNet-based network with two branches and a fusion path, as shown in Fig.~\ref{fig:DTUNet}, 
The shared layers of the DTUNet consist of three convolutional blocks, where each block has two 3D convolutional layers followed by a max pooling layer.
The first block takes a patch of $64\times64\time64\times64$ voxels as input 
and has 32 initial convolutional filters to reduce GPU memory consumption.
The number of convolutional filters in the subsequent blocks is chosen according to the design rule of 3D UNet~\cite{cciccek2016}.
The first branch is used to predict lumen segmentation and the second branch is used to predict vessel centerlines.
Both branches have the same architecture, comprising one convolutional block without the max pooling layer, three deconvolutional blocks, and one 3D convolutional layer followed by a sigmoid layer. Each deconvolution block has one 3D deconvolutional layer followed by two 3D convolutional layers. 
The fusion path (FP), which includes three fusion blocks, is positioned between the two branches to effectively aggregate the features from both branches.
The fusion block, as shown in Fig.~\ref{fig:fusion_path}, 
employs a concatenation operation and a convolutional layer to aggregate features of the two branches. It then utilizes a channel attention module and a spatial attention module
to automatically emphasize task-related features. These features are combined with the aggregated features at the end.
Specifically, given an input feature $F\in\mathbb{R}^{H\times W \times L\times C}$,
the channel attention feature $M_{c}$ is computed as:
\begin{eqnarray}
	M_{c} = F \otimes \sigma (W_{1}(W_{2}(\textrm{AvgPool}_{c}(F)+\textrm{MaxPool}_{c}(F))))
\end{eqnarray}
where $\otimes$, $\sigma$, and $W$ denote the element-wise multiplication, the sigmoid
function, and the convolutional layer, respectively. Note that $\textrm{AvgPool}_{c}$ and $\textrm{MaxPool}_{c}$
denote the average-pool and max-pool operations that are conducted along the channel dimension.
Similarly, the spatial attention feature $M_{s}$ is computed as:
\begin{eqnarray}
	M_{s} = F \otimes \sigma (W_{1}(\textrm{AvgPool}_{s}(F)+\textrm{MaxPool}_{s}(F))
\end{eqnarray}
where $\textrm{AvgPool}_{s}$ and $\textrm{MaxPool}_{s}$ denote the average-pool and max-pool operations 
that are conducted along the spatial dimension.
Given a feature of the segmentation branch $F_{s}$ and a feature of the centerline branch $F_{c}$,
the fusion block outputs a feature $F_{out}\in\mathbb{R}^{H\times W \times L\times C}$
that is formulated as:
\begin{eqnarray}
	F_{out} = \textrm{A}(F_{s}, F_{c}) \otimes (1 + M_{s}(M_{c}(\textrm{A}(F_{s}, F_{c})))
\end{eqnarray}
where $\textrm{A}$ denotes the combination of the concatenation operation and the convolutional layer.

\subsubsection{Loss function}
A topology-aware loss (TAL) function is used to optimize the DTUNet.
This TAL consists of three terms: the Dice loss for the lumen segmentation, 
the Dice loss for the vessel centerline, and the centerline Dice (clDice) loss~\cite{shit2021}.
Specifically, 
the Dice loss for the lumen segmentation is intended for the optimization of the overlap between
the predicted lumen segmentation and the ground truth of the lumen segmentation,
and is defined as:
\begin{eqnarray}
	\textrm{L}_{Dice}^{seg} = 1 - 2\frac{|S\cap S'|}{|S|+|S'|}
\end{eqnarray}
where $S$ is the ground truth of the lumen segmentation, and $S'$ is the corresponding predicted result.
The Dice loss for the vessel centerline is intended for the optimization of the overlap between
the predicted vessel centerline and the ground truth of the vessel centerline,
and is defined as:
\begin{eqnarray}
	\textrm{L}_{Dice}^{cen} = 1 - 2\frac{|C\cap C'|}{|C|+|C'|}
\end{eqnarray}
where $C$ is the ground truth of the vessel centerline, and $C'$ is the corresponding predicted result.
The clDice loss is intended for the optimization of the consistency of the topological connectivity between 
the predicted vessel centerline and the predicted lumen segmentation, and is defined as:
\begin{eqnarray}
	\textrm{L}_{clDice} = 1 - 2\frac{\textrm{Tprec}(S,C')\times \textrm{Tsens}(S',C)}{\textrm{Tprec}(S,C')+\textrm{Tsens}(S',C)}
\end{eqnarray}
where $\textrm{Tprec}(S,C')$ ensures that the predicted vessel centerline is within the ground truth
of the lumen segmentation as much as possible, 
while $\textrm{Tsens}(S',C)$ ensures that the ground truth of the vessel centerline is within
the predicted lumen segmentation. 
$\textrm{Tprec}(S,C')$ and $\textrm{Tsens}(S',C)$ are calculated as:
\begin{eqnarray}
	\textrm{Tprec}(S,C') = \frac{|S\cap C'|}{C'} \\
	\textrm{Tsens}(S',C) = \frac{|S'\cap C|}{C}
\end{eqnarray}

It is worth noting that while the $L_{clDice}$ in the TAL is identical in mathematical form to the original version of the clDice loss (i.e., soft-clDice)~\cite{shit2021}, they differ in the actual computation process.
The soft-clDice obtains the predicted centerline from the predicted lumen segmentation by using a soft-skeleton algorithm. 
In contrast, the $L_{clDice}$ directly derives the predicted centerline from the centerline branch. 
In other words, the process of computing the soft-clDice can be described as a "sequential" process, while the computation of the $L_{clDice}$ is a "parallel" process.
It means that the $L_{clDice}$ in the TAL has the potential to enhance both the lumen segmentation and the centerline extraction, whereas the original purpose of the soft-clDice was solely to improve the lumen segmentation.

For the TAL, these three loss functions are combined as follows:
\begin{eqnarray}
	\begin{split}
		\textrm{TAL} &= (1-\alpha) \cdot (0.5\cdot \textrm{L}_{Dice}^{seg} + 0.5\cdot \textrm{L}_{Dice}^{cen}) \\
		&+ \alpha \cdot \textrm{L}_{clDice} 
	\end{split}
	\label{form:TAL}
\end{eqnarray}
where $\alpha$ is a hyper-parameter that balances the loss components.

\subsection{Post-processing}
The purpose of the post-processing approaches is to get the final vessel centerlines from the  
lumen segmentation output by the segmentation branch.
As mentioned above, since the soft-clDice~\cite{shit2021} was initially used to improve the lumen segmentation, its soft-skeleton algorithm was designed to obtain the vessel centerlines that could backpropagate gradients rather than smooth the vessel centerlines.
There are many small isolated centerline segments consisting of only a few adjacent points in the vessel centerlines extracted by the soft-skeleton algorithm.
Therefore, we utilize a medial surface/axis thinning algorithm~\cite{lee1994} instead of the soft-skeleton algorithm to extract the final vessel centerline from the lumen segmentation output by the segmentation branch.
The final vessel centerlines extracted in this way are better than those directly output by the centerline branch, which is demonstrated in Section~\ref{subsec:loss_and_output}.

\section{Experiments and results}
\subsection{Data description}
The CTA images we used for our study are 
from the MR CLEAN Registry~\cite{jansen2018},
a national multi-center registry for patients 
who underwent endovascular treatment in the Netherlands between March 2014 and January 2019. 
The CTA images were routinely acquired in 17 medical centers. 
Fifty representative CTA images from patients with acute ischemic stroke were manually annotated by experienced observers~\cite{su2020} using Mevislab~\cite{heckel2009}.
Ten out of the fifty images are with intracranial vessel centerline annotations, hereinafter referred to as intracranial CTA images.
The remaining forty images have vessel centerline annotations for a randomly sampled sub-volume with $128\times 128\times128$ voxels, hereinafter referred to as cube CTA images.

We adopted a 5-fold cross-validation strategy to evaluate the proposed framework. 
In each fold, two intracranial CTA images and eight cube CTA images were used for training, 
while the remaining CTA images were used for testing.

\subsection{Implement details}
We implemented the lumen segmentation generation in MevisLab.
We implemented the pre-processing approaches, the DTUNet, and the post-processing approaches in Python and Pytorch~\cite{paszke2019}. 
We trained the DTUNet on an Nvidia GTX 2080 TI with a batch size of 4.
We applied data augmentation techniques including random flip and random elastic deformation.
We used the Adam~\cite{kingma2014} optimizer with an initial learning rate of $10^{-3}$ and a weight decay of $5\times 10^{-4}$. 

\subsection{Evaluation metrics}
We evaluated the proposed framework with average symmetric centerline distance (ASCD) and overlap (OV), following~\cite{tang2016}. These are defined as follows:

\textbf{ASCD:} given target centerline points $X$ and predicted centerline points $Y$, the average centerline distance (ACD)
is computed as:
\begin{eqnarray}
	\textrm{ACD}(X, Y) = \sum_{x\in X} min_{y\in Y} d(x, y)/|X|
\end{eqnarray}
where $d(x,y)$ is a 3D matrix consisting of the Euclidean distances between $X$ and $Y$.
And ASCD is formulated as:
\begin{eqnarray}
	\textrm{ASCD}(X, Y) = \frac{\textrm{ACD}(X,Y) + \textrm{ACD}(Y, X)}{2}
\end{eqnarray}

\textbf{OV:} 
given target centerline points $X$ and predicted centerline points $Y$, 
$\textrm{TP}^{R}_{T}$ (true positives of the reference) are those points in $X$ that have a point $y\in Y$ with a Euclidean distance less than a threshold $T$, and the other points in $X$ are $\textrm{FN}^{R}_{T}$ (false negatives of the reference).
Similarly, $\textrm{TP}^{M}_{T}$ (true positives of the model) are those points in $Y$ that have a point $x\in X$ with a Euclidean distance less than a threshold $T$, and the other points in $Y$ are $\textrm{FP}^{M}_{T}$ (false positives of the model).
In our experiments, we use two thresholds $T$ (i.e., 1.0 mm and 1.5 mm).
With these definitions, $\textrm{OV}_{T}$ is defined as:
\begin{eqnarray}
	\textrm{OV}_{T}= \frac{\textrm{TP}^{R}_{T}+\textrm{TP}^{M}_{T}}{\textrm{TP}^{R}_{T}+\textrm{TP}^{M}_{T}+\textrm{FN}^{R}_{T}+\textrm{FP}^{M}_{T}}
\end{eqnarray}

\subsection{Choosing optimization targets}

\begin{figure}[t]
	\centering
	\includegraphics[width=0.92\linewidth]{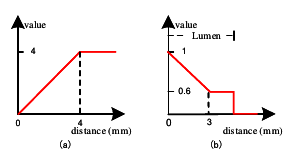}
	\caption{A diagram of (a) the distance map and (b) the heatmap.}
	\label{fig:targets}
\end{figure}

\begin{table*}[t]
	\centering
	\caption{Comparison among various types of optimization targets. $\dagger$ indicate statistically significant difference between the centerline $+$ lumen and the other optimization targets with P $<$ 0.05.}
	\scalebox{0.88}{
		\begin{tabular}{cc|ccc|ccc}
			\hline
			\multirow{2}{*}{Target} &\multirow{2}{*}{Task type} &\multicolumn{3}{c|}{Intracranial}  &\multicolumn{3}{c}{Cube}  \\
			& &ASCD (mm) $\downarrow$ &OV$_{1.0}$ $\uparrow$ &OV$_{1.5}$ $\uparrow$ &ASCD (mm) $\downarrow$ &OV$_{1.0}$ $\uparrow$ &OV$_{1.5}$ $\uparrow$ \\
			\hline
			Distance map &Regression &1.20$\pm$0.11 &0.708$\pm$0.039 &0.806$\pm$0.024 &1.78$\pm$0.46 &0.649$\pm$0.052 &0.741$\pm$0.050   \\
			Heatmap  &Regression &1.08$\pm$0.11 &0.782$\pm$0.014  &0.837$\pm$0.017 &1.49$\pm$0.40 &0.735$\pm$0.044 &0.792$\pm$0.043 \\
			Centerline &Segmentation &1.05$\pm$0.07 &0.794$\pm$0.014 &0.840$\pm$0.012 &1.66$\pm$0.37 &0.722$\pm$0.037 &0.769$\pm$0.037   \\
			Lumen &Segmentation &0.95$\pm$0.08 &0.810$\pm$0.014  &0.859$\pm$0.012 &1.40$\pm$0.39 &0.752$\pm$0.049 &0.803$\pm$0.046 \\
			Centerline $+$ Lumen &Segmentation &\textbf{0.89$\pm$0.06{$^\dagger$}} &\textbf{0.832$\pm$0.009{$^\dagger$}} &\textbf{0.880$\pm$0.009{$^\dagger$}} &\textbf{1.28$\pm$0.29{$^\dagger$}} &\textbf{0.773$\pm$0.042{$^\dagger$}} &\textbf{0.821$\pm$0.041{$^\dagger$}} \\
			\hline
	\end{tabular}}
	\label{table:ground truth}
\end{table*}

Recently, a variety of optimization targets have been adopted for vessel centerline extraction from medical images, such as
the distance map~\cite{guo2019} and the heatmap~\cite{he2020}.
In this paper, the proposed framework uses both the vessel centerline and the lumen segmentation as the optimization targets.
While the proposed framework (and similar methods) defines a segmentation task, optimizing a distance map or a heatmap is a regression task.

In order to investigate the impact of different types of optimization targets, we constructed a distance map and a heatmap based on the distance to the centerline, as shown in Fig.~\ref{fig:targets}.
The same 3D UNet was used for all four optimization targets to minimize the impact of the model. Since the proposed framework had two optimization targets (i.e., centerline + lumen), a dual-branch 3D UNet (i.e., the DTUNet without the fusion path) was used in this experiment.
Table~\ref{table:ground truth} reports the centerline extraction results of the different types of optimization targets.
Compared to the regression tasks, the lumen and the centerline + lumen both obtain better results. The centerline also obtains comparable results to the heatmap.
Moreover, in the segmentation tasks, the lumen obtains superior results to the centerline, and combining both the centerline and the lumen achieves the best results, which significantly outperforms other optimization targets.
Therefore, combining the centerline and the lumen is a well-suited optimization target for the vessel centerline extraction from CTA images.

\subsection{Hyper-parameter $\alpha$ of the topology-aware loss}
\begin{figure*}[t]
	\centering
	\includegraphics[width=0.92\linewidth]{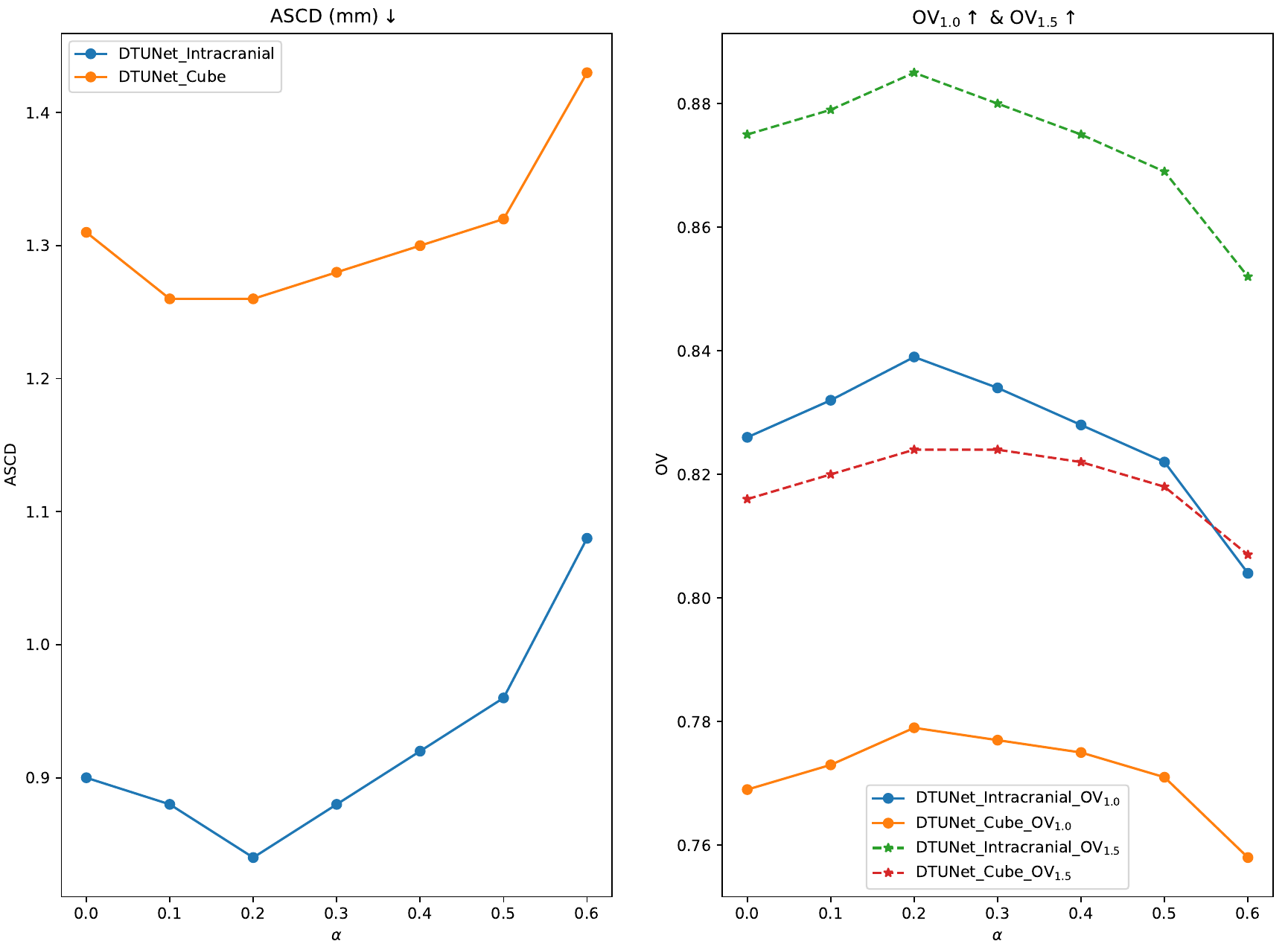}
	\caption{Results of DTUNet with different hyper-parameters $\alpha$.}
	\label{fig:coefficient}
\end{figure*}

We studied the effect of the hyper-parameter $\alpha$ of the TAL on the DTUNet.
Fig.~\ref{fig:coefficient} shows the results of the DTUNet with seven different values for $\alpha$
(i.e., 0, 0.1, 0.2, 0.3, 0.4, 0.5, 0.6).
$\alpha = 0$ means that DTUNet is optimized by $\textrm{L}_{Dice}^{seg}$ and $\textrm{L}_{Dice}^{cen}$, 
while $\alpha > 0$ means that the DTUNet is optimized by $\textrm{L}_{Dice}^{seg}$, $\textrm{L}_{Dice}^{cen}$ and $\textrm{L}_{clDice}$.
Starting from a small $\alpha$, an increase of $\alpha$ improves the performance of the DTUNet on ASCD, $\textrm{OV}_{1.0}$ and $\textrm{OV}_{1.5}$.
The DTUNet achieves the best performance at $\alpha = 0.2$; the performance of the DTUNet decreases rapidly when $\alpha >0.2$, and is even worse than the performance at $\alpha =0$ in the end.
Based on these experiments, we chose 0.2 as the value of 
the hyper-parameter $\alpha$ for DTUNet, in the following experiments unless otherwise specified.

\subsection{Topology-aware loss versus soft-clDice and soft-Dice}
\label{subsec:loss_and_output}

\begin{table*}[t]
	\centering
	\caption{Comparison between the DTUNet (without the fusion path) trained using the topology-aware loss (TAL) and the 3D UNet trained using the soft-clDice.}
	\scalebox{0.76}{
		\begin{tabular}{cc|ccc|ccc}
			\hline
			\multirow{2}{*}{Method} &\multirow{2}{*}{Output} &\multicolumn{3}{c|}{Intracranial}  &\multicolumn{3}{c}{Cube}  \\
			& &ASCD (mm) $\downarrow$ &OV$_{1.0}$ $\uparrow$ &OV$_{1.5}$ $\uparrow$ &ASCD (mm) $\downarrow$ &OV$_{1.0}$ $\uparrow$ &OV$_{1.5}$ $\uparrow$ \\
			\hline
			3D UNet + soft-clDice~\cite{shit2021}  &Lumen + thining algorithm~\cite{lee1994} &0.90$\pm$0.07 &0.825$\pm$0.008 &0.871$\pm$0.006 &1.34$\pm$0.35 &0.770$\pm$0.043 &0.815$\pm$0.042   \\
			DTUNet w/o FP + TAL &Centerline  &0.92$\pm$0.07 &0.824$\pm$0.007 &0.871$\pm$0.008 &1.33$\pm$0.34 &0.770$\pm$0.045 &0.817$\pm$0.043 \\
			DTUNet w/o FP + TAL &Lumen + thining algorithm~\cite{lee1994} &\textbf{0.89$\pm$0.06} &\textbf{0.832$\pm$0.009} &\textbf{0.880$\pm$0.009} &\textbf{1.28$\pm$0.29} &\textbf{0.773$\pm$0.042} &\textbf{0.821$\pm$0.041} \\
			\hline
	\end{tabular}}
	\label{table:loss}
\end{table*}

We compared the proposed framework with the method of Shit et al.~\cite{shit2021} (i.e., the 3D UNet trained using the soft-clDice and the soft-Dice) to demonstrate the effectiveness of the proposed TAL and its corresponding dual-branch design. 
A dual-branch 3D UNet (i.e., the DTUNet without the fusion path) was utilized as the backbone of the proposed framework to minimize the impact of the model.
As mentioned in the post-processing section, the soft-skeleton algorithm used in the soft-clDice was aimed at obtaining the vessel centerlines that could backpropagate gradients rather than smoothing the vessel centerline.
Therefore, for the 3D UNet trained using the soft-clDice and the soft-Dice, we used the medial surface/axis thinning algorithm~\cite{lee1994} to obtain the final vessel centerlines from the lumen segmentation.
Table~\ref{table:loss} reports the results of the DTUNet without fusion path (FP) trained using the TAL and the 3D UNet trained using the soft-clDice the soft-Dice.
The result of the segmentation branch of the DTUNet without the FP trained using the TAL is superior to that of the 3D UNet trained using the soft-clDice and the soft-Dice as well as that of the centerline branch.
Therefore, the proposed TAL and its corresponding dual-branch design are better suited for the vessel centerline extraction from CTA images.

\subsection{Ablation studies}

\begin{table*}[t]
	\centering
	\caption{Ablation study.}
	\scalebox{0.92}{
		\begin{tabular}{ccc|ccc|ccc}
			\hline
			\multirow{2}{*}{Backbone} &\multirow{2}{*}{Lumen} &\multirow{2}{*}{TAL} &\multicolumn{3}{c|}{Intracranial}  &\multicolumn{3}{c}{Cube}  \\
			& & &ASCD (mm) $\downarrow$ &OV$_{1.0}$ $\uparrow$ &OV$_{1.5}$ $\uparrow$ &ASCD (mm) $\downarrow$ &OV$_{1.0}$ $\uparrow$ &OV$_{1.5}$ $\uparrow$ \\
			\hline
			3D UNet & & &1.05$\pm$0.07 &0.794$\pm$0.014 &0.840$\pm$0.012 &1.66$\pm$0.37 &0.722$\pm$0.037 &0.769$\pm$0.037   \\
			3D UNet &$\surd$ &  &0.95$\pm$0.08 &0.810$\pm$0.014  &0.859$\pm$0.012 &1.40$\pm$0.39 &0.752$\pm$0.049 &0.803$\pm$0.046 \\
			DTUNet w/o FP &$\surd$ & &0.92$\pm$0.09 &0.820$\pm$0.012 &0.870$\pm$0.015 &1.33$\pm$0.41 &0.767$\pm$0.045 &0.816$\pm$0.046 \\
			DTUNet w/o FP  &$\surd$ &$\surd$ &0.89$\pm$0.06 &0.832$\pm$0.009 &0.880$\pm$0.009 &1.28$\pm$0.29 &0.773$\pm$0.042 &0.821$\pm$0.041 \\
			DTUNet   &$\surd$ & &0.90$\pm$0.08 &0.826$\pm$0.008 &0.875$\pm$0.009 &1.31$\pm$0.32 &0.769$\pm$0.043 &0.816$\pm$0.041 \\
			DTUNet  &$\surd$ &$\surd$  &\textbf{0.84$\pm$0.07} &\textbf{0.839$\pm$0.008} &\textbf{0.885$\pm$0.010} &\textbf{1.26$\pm$0.29} &\textbf{0.779$\pm$0.043} &\textbf{0.824$\pm$0.041} \\
			\hline
	\end{tabular}}
	\label{table:ablation}
\end{table*}

We also studied the impact of the lumen segmentation, the TAL, and the backbones on the performance of the DTUNet, as shown in Table~\ref{table:ablation}.
Compared to the 3D UNet that directly predicts the vessel centerline, the 3D UNet trained with the lumen segmentation
achieves statistically significantly (P $<$ 0.05) better results across all the metrics,
which means the lumen segmentation helps the 3D UNet extract the vessel centerlines from CTA images.
Table~\ref{table:ablation} also shows that the TAL has added values for both the DTUNet without the fusion path (FP) and DTUNet.
This suggests that the TAL can further improve the performance of the models by optimizing the consistency of the topological connectivity between the predicted vessel centerline and the
predicted lumen segmentation.
In addition, the DTUNet is slightly better than the DTUNet without the fusion path.
Finally, compared to the 3D UNet trained with the lumen segmentation, the DTUNet trained with the lumen segmentation obtains
statistically significant improvements (P $<$ 0.05) across all the metrics.

\begin{figure}[t]
	\centering
	\includegraphics[width=0.92\linewidth]{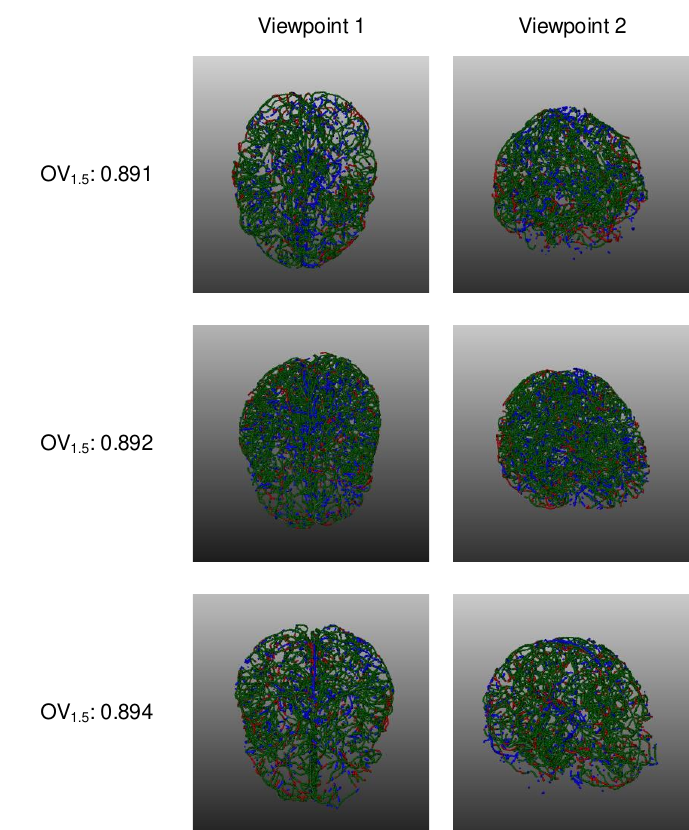}
	\caption{Visualization of three example intracranial cases with two different viewpoints. Green points are true positives; blue points are false positives; red points are false negatives. In addition, values of the OV$_{1.5}$ metric are indicated for all the cases.}
	\label{fig:vis_intra}
\end{figure}

\begin{figure}[t]
	\centering
	\includegraphics[width=0.92\linewidth]{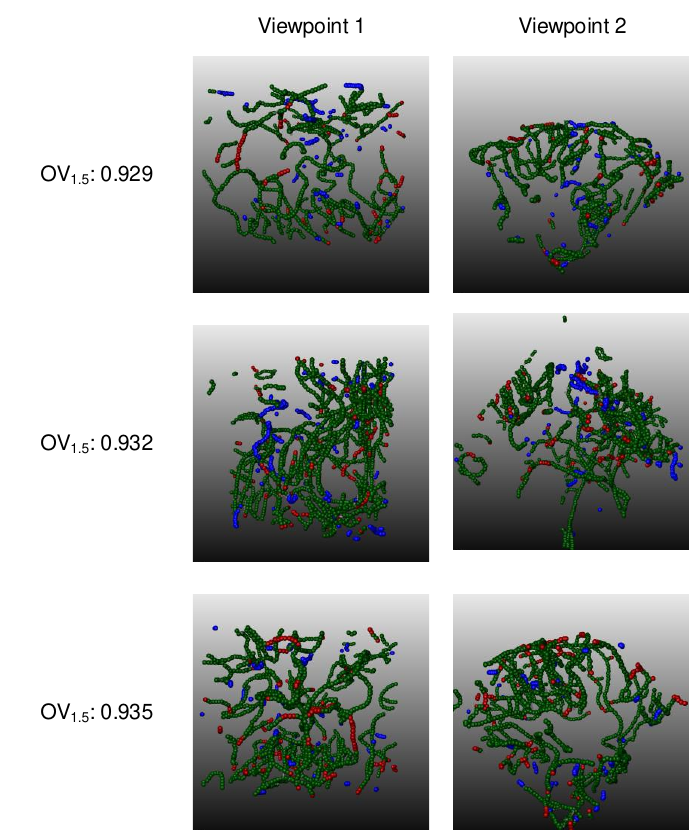}
	\caption{Visualization of three example cube cases with two different viewpoints. Green points are true positives; blue points are false positives; red points are false negatives. In addition, values of the OV$_{1.5}$ metric are indicated for all the cases.}
	\label{fig:vis_cube}
\end{figure}

\subsection{Subgroup analyses}

\begin{table}[t]
	\centering
	\caption{Subgroup analyses of the proposed framework.}
	\scalebox{0.70}{
		\begin{tabular}{cc|ccc}
			\hline
			Property &Category &ACD (mm) $\downarrow$ &Recall$_{1.0}$ $\uparrow$ &Recall$_{1.5}$ $\uparrow$ \\
			\hline
			\multirow{3}{*}{Vessel thickness} &Thin &0.87$\pm$0.11	&0.798$\pm$0.011	&0.861$\pm$0.016  \\
			&Medium &0.63$\pm$0.09	&0.875$\pm$0.017 &0.920$\pm$0.019  \\
			&Large &0.85$\pm$0.10	&0.753$\pm$0.035 &0.873$\pm$0.023  \\
			\hline
			\multirow{5}{*}{Region} &Left hemisphere &0.72$\pm$0.08	&0.847$\pm$0.016	&0.901$\pm$0.017  \\
			&Right hemisphere &0.70$\pm$0.06	&0.859$\pm$0.017	&0.910$\pm$0.018   \\
			&MCA region &0.47$\pm$0.06	&0.931$\pm$0.015	&0.963$\pm$0.017  \\
			&ACA region &0.49$\pm$0.06	&0.918$\pm$0.025	&0.958$\pm$0.019  \\
			&PCA region &0.41$\pm$0.07	&0.949$\pm$0.031	&0.983$\pm$0.018  \\
			\hline
	\end{tabular}}
	\label{table:subgroup}
\end{table}

We further computed the performance of the proposed framework on various subgroups, based on vessel radius and location.
To this end, we computed the vessel radius per segment by averaging the lumen radii of the automated segmentation.
Based on this, the segments were grouped into three categories: thin vessels (radius $\le$ 0.6mm), 
medium vessels (0.6mm $<$ radius $\le$ 1.0mm), and large vessels (radius $>$ 1.0mm).
To group the segments based on location, we applied the same registration approach mentioned in the pre-processing approaches 
to obtain a hemisphere mask, a middle cerebral artery (MCA) mask, an anterior cerebral artery (ACA) mask, and a posterior cerebral artery (PCA) mask, according to the work of~\cite{peter_cortical_2017}. 
The segments were subsequently assigned to one of these regions. 
Since we did not compute the vessel radius of the predicted vessel centerlines, we could not classify the vessel thickness types of the predicted centerlines. 
Therefore, ASCD and OV could no longer be computed, and instead we used the average centerline distance (ACD) 
that only computed the Euclidean distances from the target centerline points to the predicted centerline points, and the recall as the evaluation metrics.

Table~\ref{table:subgroup} presents the subgroup analyses of the proposed framework. It is evident from this table that the proposed framework achieves better results for medium vessels than for thin and large vessels. We observe no significant difference in the results between the left and right hemispheres, despite the brains used suffering from acute ischemic stroke.
The proposed framework also demonstrates encouraging results for the MCA region, the ACA region, and the PCA region.

\subsection{Comparison with other methods}

\begin{table*}[thb]
	\centering
	\caption{Comparison with the state-of-the-art methods. $\star$ and $\dagger$ indicate statistically significant difference between the proposed framework and the state-of-the-art methods with P $<$ 0.01 and P $<$ 0.05, respectively.}
	\scalebox{1.02}{
		\begin{tabular}{c|ccc|ccc}
			\hline
			\multirow{2}{*}{Method} &\multicolumn{3}{c|}{Intracranial}  &\multicolumn{3}{c}{Cube}  \\
			&ASCD (mm) $\downarrow$ &OV$_{1.0}$ $\uparrow$ &OV$_{1.5}$ $\uparrow$ &ASCD (mm) $\downarrow$ &OV$_{1.0}$ $\uparrow$ &OV$_{1.5}$ $\uparrow$ \\
			\hline
			3D UNet~\cite{cciccek2016}  &1.05$\pm$0.07 &0.794$\pm$0.014 &0.840$\pm$0.012 &1.66$\pm$0.37 &0.722$\pm$0.037 &0.769$\pm$0.037   \\
			VNet~\cite{milletari2016}  &1.18$\pm$0.12 &0.769$\pm$0.021  &0.820$\pm$0.019 &1.68$\pm$0.42 &0.703$\pm$0.039 &0.756$\pm$0.039 \\
			Su's net~\cite{su2020}  &1.01$\pm$0.07 &0.802$\pm$0.006 &0.849$\pm$0.007 &1.48$\pm$0.34 &0.739$\pm$0.044 &0.789$\pm$0.044 \\
			CS$^{2}$Net~\cite{mou2021} &1.05$\pm$0.10 &0.792$\pm$0.012 &0.839$\pm$0.012 &1.59$\pm$0.37 &0.720$\pm$0.036 &0.768$\pm$0.038 \\
			TransBTS~\cite{wang2021}  &0.98$\pm$0.10 &0.811$\pm$0.011 &0.856$\pm$0.017 &1.42$\pm$0.26 &0.744$\pm$0.035 &0.793$\pm$0.034 \\
			UNETR~\cite{hatamizadeh2022}  &1.18$\pm$0.08 &0.750$\pm$0.010 &0.808$\pm$0.008 &1.58$\pm$0.32 &0.694$\pm$0.043 &0.757$\pm$0.043 \\
			\hline
			Proposed   &\textbf{0.84$\pm$0.07{$^\star$}} &\textbf{0.839$\pm$0.008{$^\star$}} &\textbf{0.885$\pm$0.010{$^\star$}} &\textbf{1.26$\pm$0.29{$^\dagger$}} &\textbf{0.779$\pm$0.043{$^\dagger$}} &\textbf{0.824$\pm$0.041{$^\dagger$}} \\
			\hline
	\end{tabular}}
	\label{table:all}
\end{table*}

\begin{figure*}[t]
	\centering
	\includegraphics[width=0.92\linewidth]{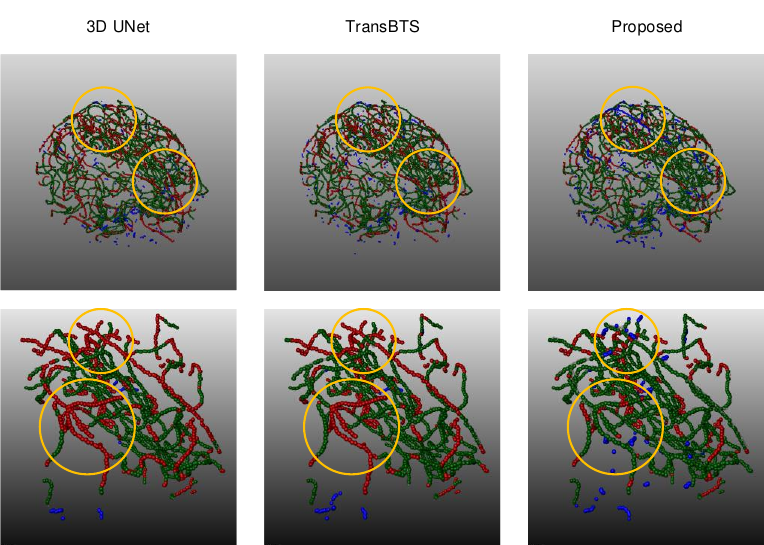}
	\caption{Visualization of the results of different methods for an intracranial case and a cube case. The first row and the second row show the results of the intracranial case and the cube case, respectively. Green points are true positives; blue points are false positives; red points are false negatives. Yellow circles highlight the areas where the results differ.}
	\label{fig:contrast}
\end{figure*}

In this experiment, we quantitatively compared the proposed framework with the state-of-the-art methods widely used for vessel segmentation and vessel centerline extraction, including 3D UNet~\cite{cciccek2016}, VNet~\cite{milletari2016},
Su's net~\cite{su2020}, CS$^{2}$Net~\cite{mou2021}, TransBTS~\cite{wang2021}, and UNETR~\cite{hatamizadeh2022}.
From Table~\ref{table:all}, it is evident that the proposed framework achieves better performance than the state-of-the-art methods across all the metrics, with statistically significant improvements.

We also qualitatively analysed the proposed framework and compared it with 3D UNet and TransBTS.
Fig.~\ref{fig:vis_intra} and Fig.~\ref{fig:vis_cube} visualize the results of the proposed framework for an intracranial case and a cube case, respectively. The proposed framework correctly predicts most of the vessel centerlines, with only a few false positives and false negatives.
Fig.~\ref{fig:contrast} visualizes the results of the proposed framework, 3D UNet, and TransBTS for an intracranial case and a cube case, respectively.
Compared with the 3D UNet and TransBTS, the proposed framework predicts more vessel centerlines correctly, at the expense of a few more false positives.

\section{Discussion}
In this paper, we presented and assessed an automated framework for brain vessel centerline extraction from CTA images.
Comprehensive experiments on a multi-center dataset demonstrated the effectiveness of each of its components and its superiority over the state-of-the-art methods.

From Table~\ref{table:ground truth}, we can find that the results of the segmentation tasks are superior to those of the regression tasks, and even the centerline achieves comparable results to the more informative heatmap.
This observation might be attributed to the effectiveness of the Dice loss used in the segmentation tasks in handling class-imbalanced samples. On the other hand, the mean square error loss employed in the regression tasks might not be well-suited for addressing the class-imbalanced issue. While more advanced loss functions may address this issue in regression tasks, we consider this beyond the scope of this paper.

As shown in Table~\ref{table:subgroup}, the proposed framework demonstrates superior results for the medium vessels than for the thin vessels and the large vessels.
This might be attributed to the fact that the centerlines of thin vessels are more likely to be missed due to their subtle nature, whereas the centerlines of large vessels are more difficult to predict as they may not be well-centered.
Besides, we observe no significant difference in the results between the left and right hemispheres in stroke patients.
This suggests that the proposed framework may be applied to the collateral circulation assessment in stroke patients. For example, the automated collateral method proposed by Su et al.~\cite{su2020} required a comparison of vascular features between an affected hemisphere and a non-affected hemisphere. 
The proposed framework also demonstrates encouraging centerline extraction performance in each of the MCA, ACA, and PCA regions. Such regions play a vital role in the treatment of stroke, such as arterial thrombolysis.
This indicates that the proposed framework holds potential in providing quantitative information on the brain vasculature, which is relevant for e.g., stroke research and treatment decision-making~\cite{peisker2017}.

While the proposed framework shows promising performance in brain vessel centerline extraction from CTA images, there is still room for improvement.
Firstly, in the regions near the brain skull, 
there are usually thinner and less intense vessels that are similar in intensity to the surrounding tissues.
They are more likely to be missed by the proposed framework since no specific constraints are adopted for such a scenario.
Therefore, more advanced network architectures focusing on thin low-contrast vessels could be further explored.
Secondly, we have observed that some of the vessel centerline segments predicted by the proposed framework are not fully connected, although these segments should belong to the same vessel.
This disconnection issue may hamper the tracking and analysis of individual vessels.
Therefore, some post-processing methods could be investigated to connect these interrupted vessel centerlines.

\section{Conclusion}
In this paper, we presented an automated framework for brain vessel centerline extraction from CTA images. 
We demonstrated that introducing the lumen segmentation generated from the annotated centerlines during the training phase did help improve the performance of the brain vessel centerline extraction without increasing annotation effort.
The proposed DTUNet, relying on its two-branch structure and topology-aware loss function, could utilize the annotated vessel centerlines and the generated lumen segmentation more effectively during the training phase, and could further achieve improved performance.
Extensive experiments on a multi-center dataset demonstrated its superiority over state-of-the-art methods on the average symmetric centerline distance (ASCD) and overlap (OV) metrics. Subgroup analyses further suggested its potential in clinical applications for stroke treatment.


\section*{Acknowledgments}
This work was supported in part by the National Key Research
and Development Program of China under grant 2017YFA0700800.\par
This work was also supported by Health-Holland (TKI Life Sciences and Health) through the Q-Maestro project under Grant EMCLSH19006 and Philips Healthcare (Best, The Netherlands).\par
We acknowledge the support of the Netherlands Cardiovascular Research Initiative which is supported by the Dutch Heart Foundation (CVON2015-01: CONTRAST), the support of the Brain Foundation Netherlands (HA2015.01.06), and the support of Health-Holland, Top Sector Life Sciences $\&$ Health (LSHM17016), Medtronic, Cerenovus and Stryker European Operations BV. The collaboration project is additionally financed by the Ministry of Economic Affairs by means of the PPP Allowance made available by the Top Sector Life Sciences $\&$ Health to stimulate public-private partnerships.\par


 

\begin{thebibliography}{00}
\bibitem{who} World Health Organization, Stroke, cerebrovascular accident.~\url{https://www.emro.who.int/health-topics/stroke-cerebrovascular-accident/index.html}, 2023, (accessed July 2023).

\bibitem{peisker2017} T. Peisker, B. Koznar, I. Stetkarova, P. Widimsky, Acute stroke therapy: A review, Trends Cardiovasc. Med. 27 (1) (2017) 59--66.

\bibitem{el-Baz2012} A. El-Baz et~al., Precise segmentation of 3-D magnetic resonance angiography, IEEE Trans. Biomed. Eng. 59 (7) (2012) 2019--2029.

\bibitem{soltanpour2021} M. Soltanpour, R. Greiner, P. Boulanger, Brian Buck, Improvement of automatic ischemic stroke lesion segmentation in CT perfusion maps using a learned deep neural network, Comput. Biol. Med 137 (2021) 104849.

\bibitem{stib2020} M. Stib et~al., Detecting large vessel occlusion at multiphase CT angiography by using a deep convolutional neural network, Radiology 297 (3) (2020) 640--649. 

\bibitem{su2020} J. Su et~al., Automatic collateral scoring from 3D CTA images, IEEE Trans. Med. Imaging 39 (6) (2020) 2190--2200.

\bibitem{lv2019} T. Lv et~al., Vessel segmentation using centerline constrained level set method, Multimed. Tools. Appl. 78 (2019) 17051--17075.

\bibitem{reinertsen2007} I. Reinertsen, F. Lindseth, G. Unsgaard,  D. Collins, Clinical validation of vessel-based registration for correction of brain-shift, Med. Image Anal. 11 (6) (2007) 673--684.

\bibitem{palagyi2001} K. Pal{\'a}gyi et~al., A sequential 3D thinning algorithm and its medical applications, in: Proc. Biennial Int. Conf. Inf. Process. Med. Imaging (IPMI), 2001, pp.~409--415.

\bibitem{cheng2014} Y. Cheng, X. Hu, Y. Wang, J. Wang,  S. Tamura, Automatic centerline detection of small three-dimensional vessel structures, J. Electron. Imaging 23 (1) (2014) 013007.

\bibitem{he2001} T. He, L. Hong, D. Chen,  Z. Liang, Reliable path for virtual endoscopy: Ensuring complete examination of human organs, IEEE Trans. Vis. Comput. graph. 7 (4) (2021) 333--342.

\bibitem{jin2016} D. Jin, K. Iyer, C. Chen, E. Hoffman,  P. Saha, A robust and efficient curve skeletonization algorithm for tree-like objects using minimum cost paths, Pattern Recognit. Lett. 76 (2016) 32--40.

\bibitem{krissian2000} K. Krissian, G. Malandain, N. Ayache, R. Vaillant,  Y. Trousset, Model-based detection of tubular structures in 3D images, Comput. Vis. Image Underst. 80 (2) (2000) 130--171.

\bibitem{fung2013} E. Fung,  R. Carson, Cerebral blood flow with [15O] water PET studies using an image-derived input function and MR-defined carotid centerlines, Phys. Med. Biol. 58 (6) (2013) 1903.

\bibitem{moccia2018} S. Moccia, E. De Momi, S. El Hadji,  L. Mattos, Blood vessel segmentation algorithms—review of methods, datasets and evaluation metrics, Comput. Methods Programs Biomed. 158 (2018) 71--91.

\bibitem{xiao2023} H. Xiao, L. Li, Q. Liu, X. Zhu, Q. Zhang, Transformers in medical image segmentation: A review, Biomed. Signal Process. Control 84 (2023) 104791.

\bibitem{guo2019} Z. Guo et~al., Deepcenterline: A multi-task fully convolutional network for centerline extraction, in: Proc. Int. Conf. Inf Process. Med. Imaging (IPMI), 2019, pp.~441--453.


\bibitem{tetteh2020} G. Tetteh el~al., Deepvesselnet: Vessel segmentation, centerline prediction, and bifurcation detection in 3-D angiographic volumes, Front. Neurosci. 14 (2020) 1285.

\bibitem{he2020} J. He et~al., Learning hybrid representations for automatic 3D vessel centerline extraction, in: Proc. Int. Conf. Med. Image Comput. Comput.-
Assis. Intervent. (MICCAI), 2020, pp.~24--34.

\bibitem{sironi2015} A. Sironi, E. T{\"u}retken, V. Lepetit,  P. Fua, Multiscale centerline detection, IEEE Trans. Pattern Anal. Mach. Intell. 38 (7) (2015) 1327--1341.

\bibitem{rjiba2020} S. Rjiba et~al., CenterlineNet: Automatic coronary artery centerline extraction for computed tomographic angiographic images using convolutional neural network architectures, in: Proc. Int. Conf. Image Process. Theory Tools Appl. (IPTA), 2020, pp.~1--6.

\bibitem{zhang2022} X. Zhang et al., X-ray coronary centerline extraction based on C-UNet and a multifactor reconnection algorithm, Comput. Methods Programs Biomed. 226 (2022) 107114.

\bibitem{zhang2018} P. Zhang, F.Wang,  Y. Zheng, Deep reinforcement learning for vessel centerline tracing in multi-modality 3D volumes, in: Proc. Int. Conf. Med. Image Comput. Comput.-Assis. Intervent. (MICCAI), 2018, pp.~755--763.

\bibitem{wolterink2019} J. Wolterink, R. van Hamersvelt, M. Viergever, T. Leiner,  I. I{\v{s}}gum, Coronary artery centerline extraction in cardiac CT angiography using a CNN-based orientation classifier, Med. Image Anal. 51 (2019) 46--60.

\bibitem{wu2020} G. Wu, L. Zhang, X. Chen, J. Lin, Y. Wang,  J. Yu, Convolutional neural network with asymmetric encoding and decoding structure for brain vessel segmentation on computed tomographic angiography, in: Proc. Int. MICCAI Brain lesion Workshop (BrainLes), 2020, pp.~51--59.

\bibitem{dorobantiu2021} A. Dorobantiu, V. Ogrean,  R. Brad, Coronary centerline extraction from CCTA using 3D-UNet, Future Internet 13 (4) (2021) 101.

\bibitem{shit2021} S. Shit et~al., clDice-a novel topology-preserving loss function for tubular structure segmentation, in: Proc. IEEE/CVF conf. Comput. Vis. Pattern Recognit. (CVPR), 2021, pp.~16560--16569.

\bibitem{pan2021} L. Pan, Z. Zhang, S. Zhang,  L. Huang, MSC-Net: Multitask learning network for retinal vessel segmentation and centerline extraction, Appl. Sci. 12 (1) (2021) 403.

\bibitem{friston2003} K. Friston, Statistical parametric mapping, in: Neuroscience databases, 2003, pp.~237--250.

\bibitem{avants2009} B. Avants, N. Tustison,  H. Johnson, Advanced normalization tools (ANTS), Insight J. 2 (365) (2009) 1--35.

\bibitem{schaap2009} M. Schaap et~al., Coronary lumen segmentation using graph cuts and robust kernel regression, in: Proc. Int. Conf. Inf Process. Med. Imaging (IPMI), 2009, pp.~528--539.

\bibitem{boykov1998} Y. Boykov, O. Veksler,  R. Zabih, Markov random fields with efficient approximations, in: Proc. IEEE conf. Comput. Vis. Pattern Recognit. (CVPR), 1998, pp.~648--655.

\bibitem{debruyne2008} M. Debruyne, M. Hubert,  J. Sukens, Model selection in kernel based regression using the influence function, J. Mach. Learn. Res. 9 (78) (2008) 2377--2400.

\bibitem{heckel2011} F. Heckel, O. Konrad, H. Hahn,  H. Peitgen, Interactive 3D medical image segmentation with energy-minimizing implicit functions, Comput. Graph. 35 (2) (2011) 275--287.

\bibitem{lee1994} T. Lee, R. Kashyap,  C. Chu, Building skeleton models via 3-D medial surface axis thinning algorithms, CVGIP Graph. Models Image Process. 56 (6) (1994) 462--478.

\bibitem{jansen2018} I. Jansen, M. Mulder,  R. Goldhoorn, Endovascular treatment for acute ischaemic stroke in routine clinical practice: prospective, observational cohort study (MR CLEAN Registry), BMJ 360 (2018).

\bibitem{heckel2009} F. Heckel, M. Schwier,  H. Peitgen, Object-oriented application development with MeVisLab and Python, Lect. Notes Inform. 154 (2009) 1338--1351.

\bibitem{paszke2019} A. Paszke et~al., Pytorch: An imperative style, high-performance deep learning library, in: Proc. Adv. Neural Inf. Process. Syst. (NeurIPS), 2019, pp.~8026--8037.

\bibitem{kingma2014} D. Kingma,  J. Ba, Adam: A method for stochastic optimization, 2014, arXiv preprint arXiv:1412.6980.

\bibitem{tang2016} S. Tang, C. Chan, Orthogonal planar search (OPS) for coronary artery centerline extraction, Signal Image Video Process. 10 (2) (2016) 335--342.

\bibitem{peter_cortical_2017} R. Peter, B. Emmer, A. van Es,  T. van Walsum, Cortical and vascular probability maps for analysis of human brain in computed tomography images, in: Proc. IEEE Int. Symp. Biomed. Imag. (ISBI), 2017, pp.~1141--1145.

\bibitem{cciccek2016} {\"O}. {\c{C}}i{\c{c}}ek, A. Abdulkadir, S. Lienkamp, T. Brox,  O. Ronneberger, 3D U-Net: Learning dense volumetric segmentation from sparse annotation, in: Proc. Int. Conf. Med. Image Comput. Comput.-Assis. Intervent. (MICCAI), 2016, pp.~424--432.

\bibitem{milletari2016} F. Milletari, N. Navab,  S. Ahmadi, V-net: Fully convolutional neural networks for volumetric medical image segmentation, in: Proc. Int Conf. 3D vis. (3DV), 2016, pp.~565--571.

\bibitem{mou2021} L. Mou et~al., CS$^{2}$-Net: Deep learning segmentation of curvilinear structures in medical imaging, Med. Image Anal. 67 (2021) 101874.

\bibitem{wang2021} W. Wang, C, Chen, M. Ding, H. Yu, S. Zha,  J. Li, TransBTS: Multimodal brain tumor segmentation using transformer, in: Proc. Int. Conf. Med. Image Comput. Comput.-Assis. Intervent. (MICCAI), 2021, pp.~109--119.

\bibitem{hatamizadeh2022} A. Hatamizadeh et~al., Unetr: Transformers for 3D medical image segmentation, in: Proc. IEEE/CVF Winter Conf. Appl. Comput. Vis. (WACV), 2022, pp.~574--584.



























\end{thebibliography}


\end{document}